\documentclass[12pt]{article}
\usepackage{enumerate}
\usepackage{ifpdf}
\usepackage{ae}
\usepackage[T1]{fontenc}
\usepackage[ansinew]{inputenc}
\usepackage{amsmath}
\usepackage{relsize}
\usepackage{amssymb}
\usepackage{tikz}
\usepackage{graphicx}
\usepackage{color}
\definecolor{darkblue}{cmyk}{0.9,0.9,0,0}
\definecolor{darkgreen}{rgb}{0,0.55,0}
\usepackage[colorlinks=true,linkcolor=darkblue,citecolor=darkblue,urlcolor=darkblue]{hyperref}
\usepackage{epsfig}
\usepackage{epstopdf}
\usepackage{graphicx}

\makeatletter
\long\def\@makecaption#1#2{
  \vskip\abovecaptionskip
  \sbox\@tempboxa{{\captionfonts #1: #2}}
  \ifdim \wd\@tempboxa >\hsize
    {\captionfonts #1: #2\par}
  \else
    \hbox to\hsize{\hfil\box\@tempboxa\hfil}
  \fi
  \vskip\belowcaptionskip}
\makeatother

\newcommand{\beq}{\begin{equation}}
\newcommand{\eeq}{\end{equation}}
\newcommand{\beqy} {\begin{eqnarray}}
\newcommand{\eeqy} {\end{eqnarray}}
\newcommand{\bsmat}{\begin{smallmatrix}}
\newcommand{\esmat}{\end{smallmatrix}}
\newcommand{\bmat}{\begin{matrix}}
\newcommand{\emat}{\end{matrix}}

\def\({\left(}
\def\){\right)}
\def\[{\left[}
\def\]{\right]}

\def\<{\langle}
\def\>{\rangle}

\usepackage{varioref}
\usepackage{makeidx}
\makeindex

\usepackage[english]{babel}
\usepackage{caption}
\usepackage{subfig}

\usepackage{tabularx}
\begin{document}

\thispagestyle{empty}

\renewcommand{\thefootnote}{\fnsymbol{footnote}}
\setcounter{page}{1}
\setcounter{footnote}{0}
\setcounter{figure}{0}
\begin{center}
$$$$

{\LARGE\textbf{\mathversion{bold}
Modular interpolating functions \\ for ${\cal N}=4$ SYM}}
\vspace{1.0cm}

\textrm{\Large Luis F. Alday and Agnese Bissi}
\\ \vspace{1.2cm}

\textit{Mathematical Institute, University of Oxford,}  \\
\textit{Radcliffe Observatory Quarter, Oxford, OX2 6GG, UK} \\
\vspace{5mm}

\par\vspace{1.5cm}

\textbf{Abstract}\vspace{2mm}
\end{center}

\noindent
We construct interpolating functions fully compatible with S-duality. We then consider the problem of resumming perturbative expansions for anomalous dimensions of low-twist low-spin non-protected operators in ${\cal N}=4$ super Yang-Mills theory. When the rank of the gauge group is small, the interpolations suggest that anomalous dimensions of leading twist operators take their maximum value at the point $\tau =\exp(i\pi/3)$. For fixed spin and large enough rank, there is a level-crossing region,  where the anomalous dimension of the leading twist operator reaches its maximum and then bounces back. 
\vspace*{\fill}

\setcounter{page}{1}
\renewcommand{\thefootnote}{\arabic{footnote}}
\setcounter{footnote}{0}

\newpage

 \def\nref#1{{(\ref{#1})}}



\section{Introduction}

${\cal N}=4$ SYM has been the subject of great attention over the last decade and substantial progress has been made. Most of the progress, however, has been restricted either to perturbation theory, the planar limit (where integrability techniques can be used) or super-symmetric/protected observables (for which localization techniques can be used).

The conformal bootstrap is a powerful tool to obtain non-perturbative information about generic conformal field theories \cite{Rattazzi:2008pe}. In \cite{Beem:2013qxa} the conformal bootstrap program was set for ${\cal N}=4$ SYM and bounds for the anomalous dimensions of leading-twist operators were obtained. These bounds are non-perturbative and planarity is not required. Furthermore, it was suggested that at certain special values of the complexified coupling constant $\tau$, these bounds are actually saturated. In \cite{Alday:2013opa} it was argued that the same occurs for structure constants (involving two protected operators and a leading-twist operator). 

A complementary way to gain access to non-perturbative physics is by re-summing the available perturbative data. These resummation techniques are particularly powerful when the answer is known at both ends. Furthermore, for theories which possess S-duality, such as ${\cal N}=4$ SYM, it is possible to introduce improved interpolating functions \cite{Sen:2013oza}. In the case of ${\cal N}=4$ SYM, S-duality implies that anomalous dimensions of leading-twist operators $\gamma(\tau)$ should be invariant under modular transformations:
\begin{equation}
\gamma(h \cdot \tau) \equiv \gamma(\frac{a \tau + b}{c \tau + d})= \gamma(\tau)
\end{equation}
where $a,b,c,d$ are integers satisfying $ad-bc=1$.  In \cite{Beem:2013hha} interpolating functions consistent with finite order subgroups of the full modular group were introduced. These interpolating functions are by construction consistent with the available perturbative data (at present up to four loops) and invariant under one of the following transformations:
\begin{equation}
h_2 \cdot \tau = -\frac{1}{\tau},~~~~~~h_3 \cdot \tau = \frac{\tau-1}{\tau}
\end{equation}
Furthermore, it is expected that these interpolating functions give a good approximation for $\gamma(\tau)$ in the whole fundamental region. According to the proposal of \cite{Beem:2013qxa}, the bounds arising from the conformal bootstrap are saturated at the duality invariant points $\tau_2=i$ and/or $\tau_3=e^{i \pi/3}$. The interpolating functions constructed in  \cite{Beem:2013hha} seem to support this proposal. 

Given the success of the above resummation techniques, it is natural to try to improve such proposal. In particular, the above interpolating functions are not invariant under the whole duality group, but just under one of the two finite order subgroups. The aim of this paper is to construct interpolating functions that are invariant under the full modular group. Our building blocks will be the real Eisenstein series which as we will see are specially tailored  for our purposes. Such interpolating functions offer several advantages, for instance, even one loop perturbative results can be resummed. In the second part of the paper we apply our methods to several cases of interest. We study the anomalous dimension of leading-twist operators and show that the interpolations suggest that the maximum values occur at $\tau=\tau_3$. Finally, we study the phenomenon of level-crossing, and argue that even for spin zero, there is a crossing region if the rank of the group is high enough ($N \gtrsim 5$).

\section{Modular interpolating functions}

Observables in ${\cal N}=4$ SYM are naturally functions of the complexified coupling constant
\begin{equation}
\tau  = y + \frac{i}{g}
\end{equation}
where $g=\frac{g_{YM}^2}{4\pi}$ and $y=\frac{\theta}{2\pi}$. ${\cal N}=4$ SYM possesses S-duality, which implies that observables should transform appropriately under modular transformations

\begin{equation}
h \cdot \tau = \frac{a \tau + b}{c \tau + d}
\end{equation}
where $a,b,c,d$ are integers satisfying $ad-bc=1$. In this note we will mostly focus on the anomalous dimensions of leading-twist operators\footnote{More precisely, superconformal primaries in long multiplets, transforming as singlets of $SU(4)$.} with spin $\ell$, which we denote $\gamma_\ell(\tau)$.  $\gamma_\ell(\tau)$ should satisfy the following properties:

\begin{itemize}
\item It should be real.
\item It should be  modular invariant.
\item Its perturbative expansion around small $g$ should contain only powers of $g$.
\end{itemize}
Our aim is to construct interpolating functions that satisfy all these properties. A natural set of building blocks are the real or non-holomorphic Eisenstein series 

\begin{equation}
E_{s}(\tau) = \frac{1}{2} \sum_{\substack{m,n \in \mathbb{Z}\\(m,n) \neq (0,0)}} \frac{Im(\tau)^s}{|m+n  \tau |^{2s}}  
\end{equation}
These are real modular invariant forms. Their small $g$ expansion is given by

\begin{equation}
E_s(\tau) =  \zeta(2s) \frac{1}{g^s}+\frac{\pi^{1/2}\Gamma(s-1/2)}{\Gamma(s)} \zeta(2s-1) g^{s-1}+ f^{np}_s(q),~~~~s>1
\end{equation}
where $ f^{np}_s(q)$ denotes non-perturbative contributions ({\it i.e.} they are identically zero in perturbation theory) containing powers of $q=e^{2\pi i \tau} = e^{2\pi i y} e^{-2\pi\frac{1}{g}}$. $E_{s}(\tau)$ develops a pole at $s=1$. After substracting this pole we obtain 

\begin{equation}
\lim_{s->1}\left(E_s(\tau)- \frac{\pi/2}{s-1} \right)=\zeta(2)\frac{1}{g} +\frac{\pi}{2} \log g+C-\pi \sum_{r=1}^\infty \log|1-q^r|^2
\end{equation}
For some constant $C$ not important for our purposes. Given a perturbative expansion of the form

\begin{equation}
\gamma(g)= \alpha_1 g + \alpha_2 g^2 +...+\alpha_{m} g^{m} +... 
\end{equation}
the small $g$ expansion of the real Eisenstein series suggests an interpolating function of the form
\begin{equation}
\gamma^{int,m}(\tau) = \left( c_{m+1} E_{m+1} +c_{m} E_{m} +...+c_2 E_2 \right)^{-\frac{1}{m+1}}
\end{equation}
where the coefficients $c_i$ are fixed by requiring the interpolating function to reproduce the perturbative data up to $m$ loops. This family of interpolating functions has nice features:

\begin{itemize}
\item For any $m$, the function is truly invariant under the full modular group.
\item Interpolating functions consistent with a finite order subgroup are expected to be most accurate around the corresponding invariant point. For that reason, the above interpolating functions are expected to be quite accurate it the whole fundamental region.
\item The perturbative expansion of $\gamma^{int}(\tau)$ contains only powers of $g$.
\item The interpolating functions work for any number of loops, even one-loop.
\item There are no continuous parameters to play with.\footnote{However, see the appendix A for generalizations.
}
\end{itemize}
In the following section we will apply these interpolating functions to several quantities of interest.

\section{Applications}

\subsection{Anomalous dimensions}

The most natural application of the above interpolating functions are anomalous dimensions of leading twist-operators. There are available perturbative results for any spin up to three loops and for spin zero and two up to four loops.\footnote{For the present purposes the full, non-planar, result is important.} Given a general expansion 
\begin{equation}
\gamma = \alpha_1 g +\alpha_2 g^2+\alpha_3 g^3+\alpha_4 g^4+...
\end{equation}
The corresponding interpolating functions to one, two, three and four loops are
\begin{eqnarray}
\gamma^{int,1} &=& \alpha_1 \left(\frac{E_2}{\zeta(4)} \right)^{-1/2} \\
\gamma^{int,2} &=& \left(\frac{1}{\zeta(6) \alpha_1^3} E_3-\frac{3\alpha_2}{\zeta(4) \alpha_1^4} E_2 \right)^{-1/3} \\
\gamma^{int,3} &=& \left(\frac{1}{\zeta(8) \alpha_1^4} E_4 - \frac{4\alpha_2}{\zeta(6) \alpha_1^5} E_3 + \frac{2(5 \alpha_2^2-2\alpha_1 \alpha_3)}{\zeta(4)\alpha_1^6} E_2 \right)^{-1/4} \\
\gamma^{int,4}&=&\left( \frac{1}{\zeta(10) \alpha_1^5} E_5- \frac{5\alpha_2}{\zeta(8) \alpha_1^6} E_4+ 5\frac{3 \alpha_2^2-\alpha_1 \alpha_3}{\zeta(6) \alpha_1^7} E_3-5\frac{7\alpha_2^3-6\alpha_1\alpha_2\alpha_3+\alpha_1^2\alpha_4}{\zeta(4)\alpha_1^8} E_2 \right)^{-1/5}
\end{eqnarray}
Let us apply the method to the anomalous dimensions of spin zero and two. Up to four loops, those are given by \cite{Velizhanin:2008jd}
\begin{eqnarray}
\gamma_0(g)&=& \frac{3 N g}{\pi}- \frac{3N^2 g^2}{\pi^2}+\frac{21 N^3 g^3}{4\pi^3}+ (-39+9 \zeta(3)-45 \zeta(5)(\frac{1}{2} + \frac{6}{N^2})) \frac{N^4 g^4}{4\pi^4} +...\\
\gamma_2(g)&=& \frac{25 N g}{6 \pi} - \frac{925 N^2 g^2}{216 \pi^2}+ \frac{241325 N^3 g^3}{31104 \pi^3} +\\
&+& \left(-\frac{8045275}{2187}+\frac{114500}{81}\zeta(3)- \frac{25000}{9} \zeta(5)  +\frac{8400+28000 \zeta(3)-100000 \zeta(5)}{3 N^2}\right)\frac{N^4 g^4}{(4\pi)^4} +...\nonumber
\end{eqnarray}
The following figures show the interpolating functions for $N=2$ and $m=1,2,3,4$.

\begin{figure}[h]
\begin{center}
\includegraphics[width=6.7in]{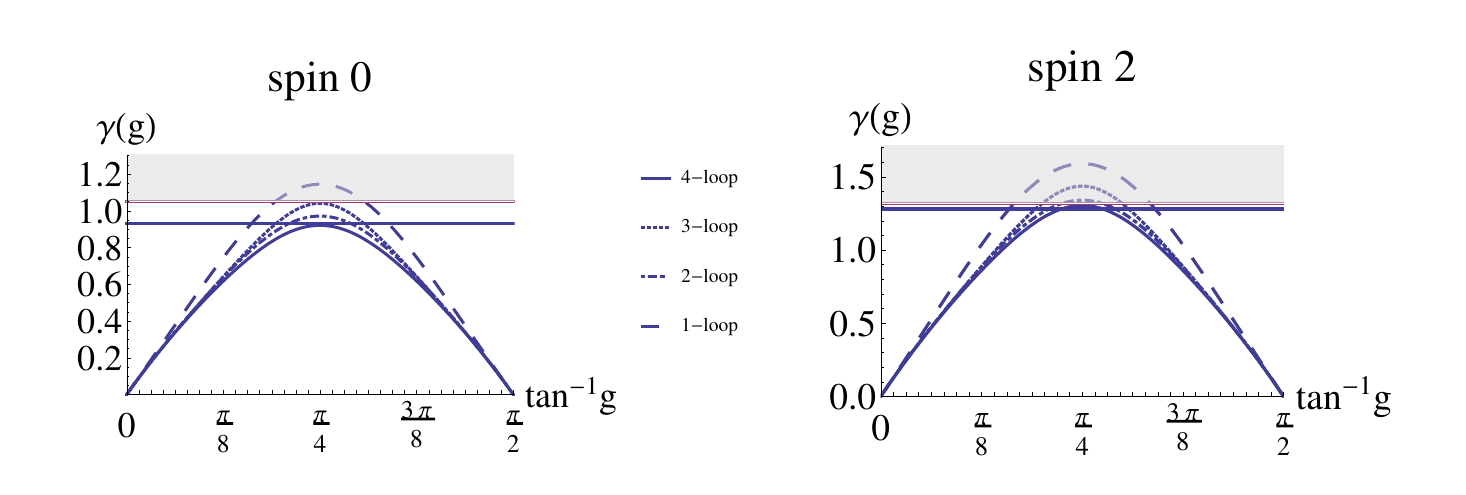}
\end{center}
\caption{Interpolating functions for anomalous dimensions of leading-twist operators with spin zero and two, for $SU(2)$ and $m=1,2,3,4$. In both cases, the upper line corresponds to $m=1$, the second line to $m=3$, the third line to $m=2$ and the lower line to $m=4$. We have also indicated the forbidden region with gray and the corner value with a continuous horizontal line.}
\end{figure}

From these figures we can draw several conclusions. First, we see that as we increase the number of loops taken into account (the label $m$), the interpolating functions seem to converge. Furthermore, even for $m=1$, where only the one-loop result has been resummed, we get a pretty accurate interpolating function. 

Having fixed the interpolating functions from the perturbative data, a natural question is what is their value at the duality invariant points $\tau=\tau_2=i$ and $\tau=\tau_3=e^{i\pi/3}$. At those values the building blocks become
\begin{eqnarray}
E_{s}(i) &=& \frac{1}{2} \sum_{\substack{m,n \in \mathbb{Z}\\(m,n) \neq (0,0)}} \frac{1}{(m^2+n^2)^s} = \frac{2 \zeta(s)}{4^s}\left( \zeta_{1/4}(s)-\zeta_{3/4}(s) \right) \\
E_{s}(e^{i\pi/3}) &=& \frac{1}{2} \frac{3^{s/2}}{2^s} \sum_{\substack{m,n \in \mathbb{Z}\\(m,n) \neq (0,0)}} \frac{1}{(m^2+m n+n^2)^s}  = \frac{1}{2^s 3^{s/2-1}} \zeta(s) \left( \zeta_{1/3}(s)-\zeta_{2/3}(s)\right)
\end{eqnarray}
where $\zeta_{a}(s)$ denotes the Hurwitz zeta function.The following tables show the results for $\gamma^{int,4}(\tau_2)$ and $\gamma^{int,4}(\tau_3)$ for operators with spin zero and two and several values of $N$.

\begin{center}
\begin{table}[htdp]
\hspace{0.5cm}
\begin{minipage}[b]{0.45\linewidth}\centering
\begin{tabular}{|c|c|c|c|}\hline  & $\gamma_{int}(i)$ & $\gamma_{int}(e^{i\pi/3})$ & $\gamma_{corner}$ \\\hline $SU(2)$ & 0.92 & 0.94 & 0.93 \\\hline $SU(3)$ & 1.238 & 1.26 &1.24 \\\hline $SU(4)$ & 1.51 & 1.536 & 1.47 \\\hline \end{tabular} \caption{Interpolating function at duality invariant points $\tau_2$ and $\tau_3$ for spin-zero, together with the corner values.}
\end{minipage}
\hspace{0.5cm}
\begin{minipage}[b]{0.45\linewidth}
\begin{tabular}{|c|c|c|c|}\hline  & $\gamma_{int}(i)$& $\gamma_{int}(e^{i\pi/3})$  & $\gamma_{corner}$ \\\hline $SU(2)$ & 1.30 & 1.33 & 1.28 \\\hline $SU(3)$ & 1.75 & 1.79 & 1.6 \\\hline $SU(4)$ & 2.13 & 2.18 & 1.75 \\\hline \end{tabular} \caption{Interpolating function at duality invariant points $\tau_2$ and $\tau_3$ for spin-two, together with the corner values.}
\end{minipage}
\end{table}
\end{center}
In the tables we have also included the corner values $\gamma_{corner}$. According to the conjecture in \cite{Beem:2013qxa} these should be the values of the anomalous dimensions at one (or both) of the duality invariant points. We see that in the low rank/lower spin cases, the value at the duality points coincides surprisingly well with the corner values! (this can also be seen explicitly in fig. 1 for $SU(2)$). On the other hand,  $\gamma_{int}(\tau_2)$ and $\gamma_{int}(\tau_3)$ become higher than $\gamma_{corner}$ as we increase the rank of the group and/or the spin. We give an interpretation of this fact at the end of this section.

Finally, we note the following interesting point. For all cases we have analyzed, the coefficients $c_k$ of the interpolation  $\left( c_m E_m +...+c_2 E_2 \right)^{-\frac{1}{m}}$ are always positive. On the other hand we can study $E_k(\tau)$ on the whole fundamental region and ask where does it take its minimal value. It turns out this happens, for all $k$, at $\tau = \tau_3$. Both facts together suggest the following:

\bigskip

{\it
\noindent The anomalous dimensions of leading twist-operators in ${\cal N}=4$~SYM take their maximum value at $\tau=e^{i\pi/3}$. It is at this value of the coupling constant that we expect the bounds arising from the conformal bootstrap to be saturated.
}

\bigskip

In appendix A we show that this is still true for a large family of modular invariant interpolating functions. 

\subsection{Structure constants}

Another quantity of interest that can be studied with the interpolating method presented in this paper are structure constants. Structure constants have a perturbative expansion of the form

\begin{equation}
a(g) = \beta_0 - \beta_1 g+\beta_2 g^2 +...
\end{equation}
where we have stressed the fact that the term proportional to $g$ is negative for the cases we will consider. The structure constants of interest are known up to three loops. After substracting the tree level contribution, we can apply the method above. The interpolating functions up to three loops are given by
\begin{eqnarray}
a^{int,1}(\tau) &=& \beta_0 -\beta_1 \left(\frac{E_2}{\zeta(4)} \right)^{-1/2} \\
a^{int,2}(\tau) &=& \beta_0 - \left( \frac{1}{\zeta(6) \beta_1^3} E_3+ \frac{3\beta_2}{\zeta(4) \beta_1^4}E_2 \right)^{-1/3}\\
a^{int,3}(\tau) &=& \beta_0 - \left(\frac{1}{\zeta(8) \beta_1^4} E_4 + \frac{4\beta_2}{\zeta(6) \beta_1^5} E_3 + \frac{2(5 \beta_2^2-2\beta_1 \beta_3)}{\zeta(4)\beta_1^6} E_2 \right)^{-1/4}
\end{eqnarray}
Using the explicit perturbative results for spin zero and spin two  structure constants/OPE coefficients we can build their interpolating functions. The following figure shows the results for one, two and three-loop.

\begin{figure}[h]
\begin{center}
\includegraphics[width=6.7in]{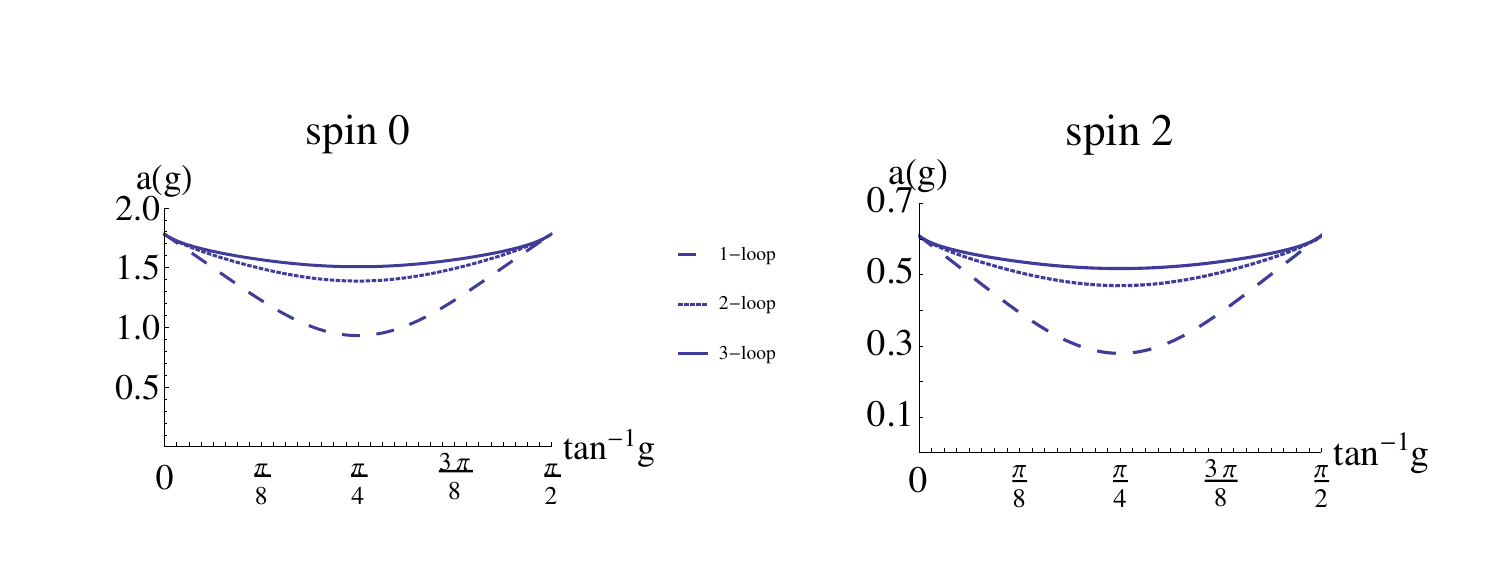}
\end{center}
\caption{Interpolating functions for structure constants of operators with spin zero and two, for $SU(2)$ and $m=1,2,3$. The lower line corresponds to $m=1$, the line in the middle to $m=2$ and the upper line to $m=3$.}
\end{figure}
The interpolating functions appear to converge as we increase the value of $m$. Furthermore, for the case of spin zero, the interpolating functions at $g=1$ seem to approach the bound value $1.6$, found in \cite{Alday:2013opa}. On the other hand, the bound found for spin two (around $0.8$) seems higher than the interpolating results.

\subsection{Level-crossing}

In the following we will study the phenomenon of level-crossing. Consider the leading-twist operator with spin $\ell$. At weak coupling its dimension is given by
\begin{equation}
\Delta_\ell^{l} = 2 + \ell + 2 h(\ell+2) \frac{N g}{\pi}+...
\end{equation}
where $h(\ell+2)$ is the harmonic number. We can also consider operators with the same spin and sub-leading twist, which at tree-level have dimension $\Delta=\ell+4$. Already at spin zero there is a degeneracy of such operators. This degeneracy is broken as we turn on the coupling and we can focus on the operator with the smallest dimension. At weak coupling its dimension is given by
\begin{equation}
\Delta_\ell^{sl} = 4 + \ell - c(N,\ell) g+...
\end{equation}
where $c(N,\ell)$ is some constant that depends on the spin and the rank of the gauge group (not necessarily large). In particular, we see that $\Delta_\ell^{l}$ grows as we increase the coupling, while $\Delta_\ell^{sl}$ decreases. We would like to understand whether it is possible that at some value of the coupling  $\Delta_\ell^{l} $ actually reaches $\Delta_\ell^{sl}$. At this point two things could happen. Additional symmetries (such as integrability) may forbid mixing between the operators, in which case the dimensions will actually cross-over. Otherwise, there will be a (small) mixing of operators and the dimensions of the new eigenstates will repel, according to the Wigner - von Neumann theorem. Even though there is no actual crossing in the second case, we will refer to both situations as level crossing. Note that in both cases, the dimension of the real leading-twist operator is the lowest one. 

We expect level-crossing to happen for large enough values of the spin. When the spin is large, the anomalous dimension of leading twist operators grows logarithmically as
\begin{equation}
\Delta_\ell^{l} = 2 + \ell + f(g,N) \log \ell +...
\end{equation}
On the other hand, for sufficiently large spin we have \cite{Alday:2007mf} \footnote{The precise expression is $\Delta_\ell^{sl} = 2 \Delta_0^{l} - \frac{d(N)}{s^2} +...$ but $\Delta_0^{l}$ does not depend on the spin and is bounded.}
\begin{equation}
\Delta_\ell^{sl} \approx 4+\ell - \frac{d(N)}{s^2} +...
\end{equation}
hence, at some value of the coupling constant, of order $g \sim (N \log \ell)^{-1}$, the dimensions of $\Delta_\ell^{l}$ and $\Delta_\ell^{sl}$ will cross over.

In the following we use interpolating functions to study level-crossing for operators with spin zero. The leading-twist operator is the Konishi operator, of the form $Tr \Phi^I \Phi^I $, whose anomalous dimension up to four loops was given in the previous section. There are four operators with twist four and spin zero:
\begin{eqnarray}
\mbox{Tr} \, \Phi^I \Phi^I \Phi^J \Phi^J,~~~~~\mbox{Tr} \, \Phi^I \Phi^J \Phi^I \Phi^J,~~~~~\mbox{Tr} \, \Phi^I \Phi^I  \mbox{Tr} \,  \Phi^J \Phi^J,~~~~~ \mbox{Tr} \,  \Phi^I \Phi^J  \mbox{Tr} \,  \Phi^I \Phi^J
\end{eqnarray}
the eigenvectors of the dilatation operator are actually specific linear combinations of those. Their anomalous dimension at one loop is given by \cite{Arutyunov:2002rs,Beisert:2003tq}

\begin{equation}
\Delta_0^{sl}(g) = 4 + \omega(N)  \frac{N}{2\pi} g+...
\end{equation}
where $\omega$ is one of the roots of 
\begin{equation}
\label{twistfourdim}
\omega^4-25 \omega^3+\left(188-\frac{160}{N^2} \right)\omega^2 - \left(384-\frac{1760}{N^2} \right)\omega - \frac{7680}{N^2}=0
\end{equation}
For the special case $N=2$ , eq. (\ref{twistfourdim}) has only two real solutions. This corresponds to the fact that in this case $\Phi^I$ are $SU(2)$ matrices, and only two of the states listed above are independent. For each value of $N$ we can solve (\ref{twistfourdim}). The smallest root is always negative and we call it $\omega_-$. Then, the one-loop interpolating function for the sub-leading-twist operator is

\begin{equation}
\Delta_0^{sl,int}(\tau)= 4 - \frac{N \omega_-}{2\pi} \sqrt{\frac{\zeta(4)}{E_2(\tau)}}
\end{equation}
While the interpolating function for the leading twist operator was constructed above.The following plots show $\Delta_0^{l,int}(\tau)$ vs $\Delta_0^{sl,int}(\tau)$ for several values of $N$.
\begin{figure}[h]
\begin{center}
\includegraphics[width=6.7in]{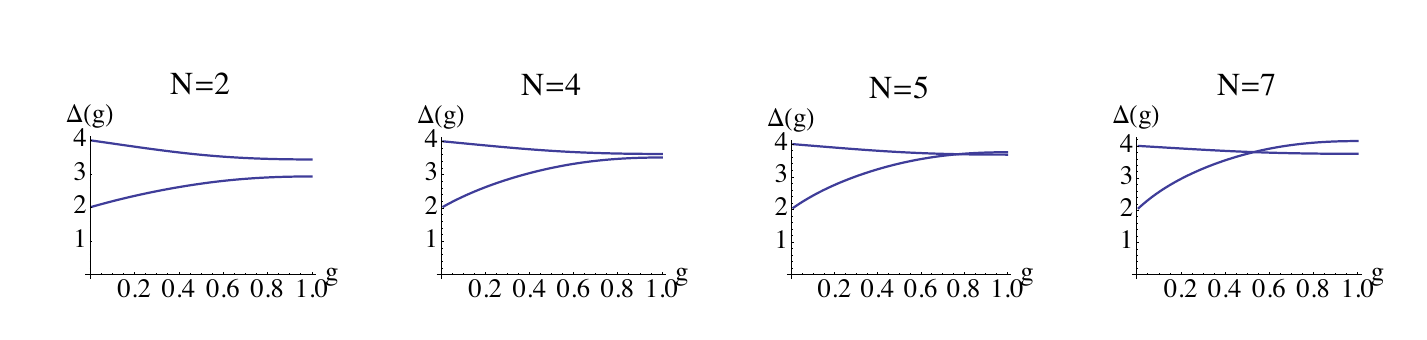}
\end{center}
\caption{Dimensions of leading and sub-leading operators of spin zero, for several values of $N$. The interpolating functions actually suggest that crossing does occur for $N$ bigger than four/five.}
\end{figure}
The interpolating functions actually suggest that crossing does occur for $N$ bigger than four. This of course could slightly change if two-loop results for the anomalous dimension of twist-four operators become available, but we expect crossing will still be present. As we increase the spin, we expect level-crossing to occur at smaller and smaller values of the rank $N$. Unfortunately, to the best of our knowledge, there are no available results (exact in $N$) for the anomalous dimension of twist-four operators with $\ell=2,4,...$. \footnote{Precisely we are interested in results exact in $N$ for superconformal primaries in long multiplets. In particular, these states are singlets of $SU(4)$, which makes the mixing problem quite involved.}

Note that level crossing would explain why for sufficiently large $N$ and/or $\ell$ the value of the interpolating functions at $\tau=\tau_2,\tau_3$ is always (quite) larger than the corner values \footnote{In addition one would expect the interpolating functions to be less accurate for large values of $N$, since the effective coupling constant is $N g$, but this by itself doesn't seem to explain why the disagreement is always in excess.}. For sufficiently large $N$ the maximum value of the anomalous dimension of the leading-twist operator is the value of $\gamma$ at the crossing region. 

For finite $N$ we expect the Wigner - von Neumann rule to apply, and hence there won't be actually crossing (as mentioned above). In this case the interpolating functions are reliable up to the crossing region. It would be interesting to understand precisely what happens at the crossing region and afterwards. \footnote{We thank Slava Rychkov for discussions on this point.}

\section{Discussion}

In this paper we have constructed interpolating functions invariant under the full modular group $PSL(2,\mathbb{Z})$. These functions can be used to study any modular invariant observables and present several nice features. In particular we have studied anomalous dimensions and structure constants of low-twist/low-spin operators in ${\cal N}=4$ SYM. 

When applied to anomalous dimensions of leading-twist operators the above interpolating functions seem to strongly suggest that the anomalous dimensions will take their maximum value at $\tau=\tau_3$. Following \cite{Beem:2013qxa} it is then natural to propose that at $\tau=\tau_3$ the bounds from the conformal bootstrap are actually saturated (at least for low enough rank/spin). We have also studied the phenomenon of level-crossing and showed that even for spin zero, level-crossing can happen when the rank of the gauge group is higher than four/five. 

There are several open problems that would be interesting to address.  First, there is much we can do if we had at our disposal further perturbative results (exact in $N$ and for low spin). {\it e.g.} five loop results for the Konishi operator would give better estimates and would allow to test whether the coefficients in the interpolating functions are positive. This is true up to four loops, and seems to be true to very high order in the planar limit. Two-loop results for twist-four spin zero operators and/or one loop for higher spin, would allow to study the level crossing phenomenon with higher precision and in more generality. 

The results of this paper suggest a refined version of the conjecture in  \cite{Beem:2013qxa}, namely the bounds from the conformal bootstrap are saturated at $\tau=\tau_3$. It would be interesting to understand the implications of this fact. A fascinating possibility would be the existence of extra symmetries at this point. This would imply additional structure for the spectrum at $\tau=\tau_3$. If further perturbative data becomes available, we could explore this possibility with the use of interpolating functions.

It was observed in \cite{Beem:2013hha} that the ratios of anomalous dimensions for operators with different spins were approximately constant over the whole fundamental region. The interpolating functions constructed in this paper imply
$$\frac{\gamma_\ell^{int,m}(\tau)}{\gamma_0^{int,m}(\tau)} = \left( \frac{\gamma_2}{\gamma_0}\right)_{tree}  \left( R^{\gamma}_\ell(\tau) \right)^{-\frac{1}{m}}$$
For $R^{\gamma}_\ell(\tau)$ some rational function of real Eisenstein series. Note that this function is one at tree-level, and at generic values of the coupling the power $1/m$ brings it closer to one. Still, $R^{\gamma}_\ell(\tau)$ itself seems to have a very mild dependence on the coupling. It would be interesting to understand the reason for this. 

It would also be very interesting to guess an exact modular invariant expression for the anomalous dimension of the Konishi operator, {\it e.g.} for $SU(2)$. More generally, one can ask whether two functions with the same perturbative expansion can differ non-perturbatively (requiring real, modular invariant functions). When using interpolating functions, we are somehow assuming that this does not happen, but it would nice to understand this issue in more detail.  Finally, it would be interesting to apply the modular invariant interpolating functions to other contexts, such as \cite{Sen:2013oza,Pius:2013tla}.

\subsection*{Acknowledgments}
\noindent
We would like to thank Philip Candelas, Slava Rychkov and Ashoke Sen for enlightening discussions. The work of the authors is supported by ERC STG grant 306260. L.F.A. is a Wolfson Royal Society Merit Award holder.

\appendix

\section{Alternative interpolating functions}

In this appendix we explore a family of modular invariant interpolating functions that generalize the ones introduced in the body of the paper. These take the form
\begin{eqnarray}
\gamma(\tau)^{n,L}_{int}= \left( c_n E_n +c_{n+1} E_{n+1} +...+c_{n+L-1} E_{n+L-1} \right)^{-\frac{1}{n+L-1}}
\end{eqnarray}
where $n$ is not necessarily an integer. The coefficients $c_n,...,c_{n+L-1}$ can be fixed by matching the $L-$loop perturbative result. Since $E_n$ develops a pole at $n=1$, we choose $n>1$. The case studied in the body of the paper corresponds to $n=2$, which is the minimal case with integer coefficients.  Another natural choice would be $n=3/2$. 

We have applied the above interpolating functions to the anomalous dimensions of leading-twist operators with spin zero and two. For instance, the following plots show the results for $N=2$ and $n=3/2,2,5/2$.

\begin{figure}[h]
\begin{center}
\includegraphics[width=5in]{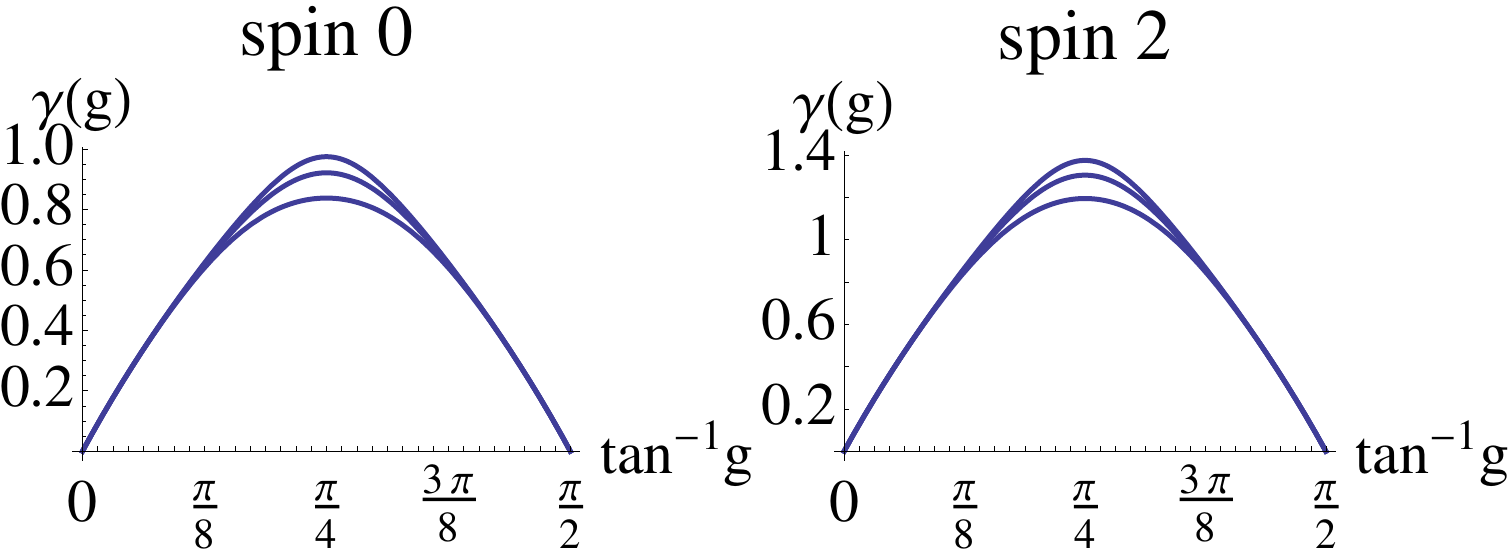}
\end{center}
\caption{Interpolating functions for $n=3/2$ (lower curve), $n=2$ (middle curve) and $n=5/2$ (upper curve), for spin zero and two.}
\end{figure}

From studying the above interpolating functions for several values of $n$ we can draw the following conclusions. 

\begin{itemize}
\item For $n$ close to two, the curves do not differ much. In all cases the maximum of the curves increases as we increase $n$. Interestingly, the corner value at $\tau=\tau_2$ and/or $\tau=\tau_3$ is best approximated by $n=2$, which seems also the most natural choice. 
\item For all cases the coefficients in the interpolating functions are positive. Hence, for the whole family of modular interpolating functions the anomalous dimensions take their maximum value at the duality invariant point $\tau=\tau_3$, which show that this result is quite robust.
\end{itemize}

\section{A toy model}

Let us apply the procedure sketched in the body of the paper to the following toy model

\begin{equation}
\gamma_{toy-pert}(g) = \sum_\ell (-1)^{\ell+1} \alpha^\ell g^\ell 
\end{equation}
this is a very simple toy model, but has some of the features of the anomalous dimension of the Konishi operator in the planar limit, namely, is an alternating sum with a finite radius of convergence, see for instance \cite{Leurent:2013mr}. Of course, we could perform the above sum:

\begin{equation}
\gamma_{toy-pert}(g) = \frac{\alpha g}{1+\alpha g}
\end{equation}
but the answer is not modular invariant. The full modular invariant answer should also contain non-perturbative corrections. It turns our that one can compute the interpolating functions exactly, for every value of $m$, obtaining
\begin{equation}
\gamma_{toy}^{int,m}(\tau) = \left( \sum_{k=2}^{m+1} \frac{(m+1)!}{k! (m+1-k)!} \frac{1}{\alpha^k \zeta(2k)} E_k(\tau) \right)^{-\frac{1}{m+1}}
\end{equation}
We have studied the above functions numerically for larger and larger values of $m$. In the limit, the functions $\gamma_{toy}^{int,m}(\tau)$ converge to a modular invariant function $\gamma_{toy}^{int,\infty}(\tau)$ which coincides with $\gamma_{toy-pert}(g)$ in the full region $0 \leq g \leq 1$, but with the correct modular properties! see figure

\begin{figure}[h]
\begin{center}
\includegraphics[width=3in]{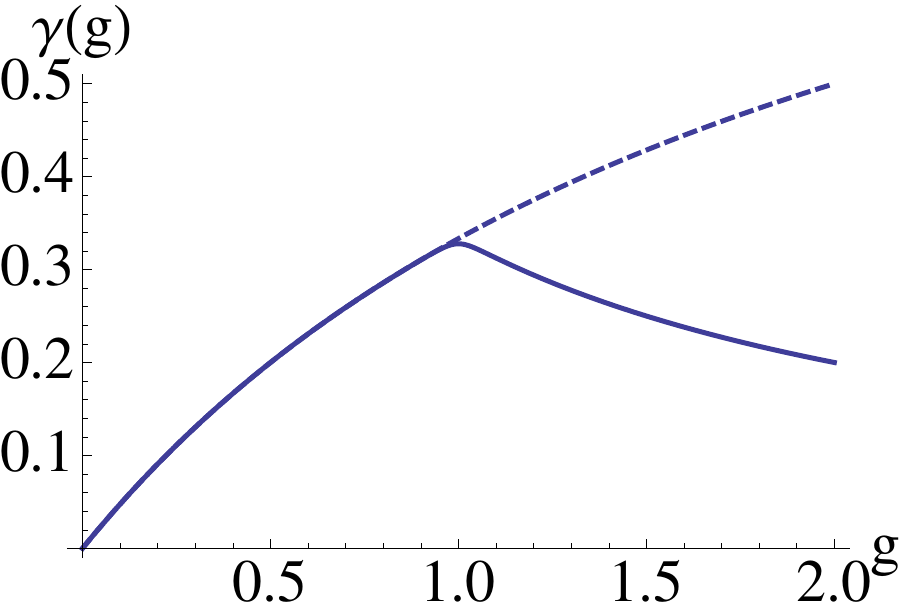}
\end{center}
\caption{$\gamma_{toy-pert}(g)$ for $\alpha=1/2$ (dashed line) vs the modular invariant interpolating function (solid line) for $m=40$.}
\end{figure}

Finally, note that all the coefficients in $\gamma_{toy}^{int,m}(\tau)$ are positive. This is also the case for the precise planar result of the Konishi operator.



\begin{thebibliography}{99}

\bibitem{Rattazzi:2008pe}
  R.~Rattazzi, V.~S.~Rychkov, E.~Tonni and A.~Vichi,
  ``Bounding scalar operator dimensions in 4D CFT,''
  JHEP {\bf 0812} (2008) 031
  [arXiv:0807.0004 [hep-th]].

\bibitem{Beem:2013qxa}
  C.~Beem, L.~Rastelli and B.~C.~van Rees,
  ``The N=4 Superconformal Bootstrap,''
  Phys.\ Rev.\ Lett.\  {\bf 111} (2013) 071601
  [arXiv:1304.1803 [hep-th]].
  

\bibitem{Alday:2013opa}
  L.~F.~Alday and A.~Bissi,
  ``The superconformal bootstrap for structure constants,''
  arXiv:1310.3757 [hep-th].
  
\bibitem{Sen:2013oza}
  A.~Sen,
  ``S-duality Improved Superstring Perturbation Theory,''
  arXiv:1304.0458 [hep-th].
  
\bibitem{Beem:2013hha}
  C.~Beem, L.~Rastelli, A.~Sen and B.~C.~van Rees,
  ``Resummation and S-duality in N=4 SYM,''
  arXiv:1306.3228 [hep-th].
  
  
\bibitem{Velizhanin:2008jd}
  V.~N.~Velizhanin,
  ``The four-loop anomalous dimension of the Konishi operator in N=4 supersymmetric Yang-Mills theory,''
  JETP Lett.\  {\bf 89} (2009) 6
  [arXiv:0808.3832 [hep-th]].
  
  V.~N.~Velizhanin,
  ``The Non-planar contribution to the four-loop anomalous dimension of twist-2 operators: First moments in N=4 SYM and non-singlet QCD,''
  Nucl.\ Phys.\ B {\bf 846} (2011) 137
  [arXiv:1008.2752 [hep-th]].
  

  
\bibitem{Alday:2007mf}
  L.~F.~Alday and J.~M.~Maldacena,
  ``Comments on operators with large spin,''
  JHEP {\bf 0711} (2007) 019
  [arXiv:0708.0672 [hep-th]].
  
\bibitem{Arutyunov:2002rs}
  G.~Arutyunov, S.~Penati, A.~C.~Petkou, A.~Santambrogio and E.~Sokatchev,
  ``Nonprotected operators in N=4 SYM and multiparticle states of AdS(5) SUGRA,''
  Nucl.\ Phys.\ B {\bf 643} (2002) 49
  [hep-th/0206020].

\bibitem{Beisert:2003tq}
  N.~Beisert, C.~Kristjansen and M.~Staudacher,
  ``The Dilatation operator of conformal N=4 superYang-Mills theory,''
  Nucl.\ Phys.\ B {\bf 664} (2003) 131
  [hep-th/0303060].
  
\bibitem{Leurent:2013mr}
  S.~Leurent and D.~Volin,
  ``Multiple zeta functions and double wrapping in planar $N=4$ SYM,''
  Nucl.\ Phys.\ B {\bf 875} (2013) 757
  [arXiv:1302.1135 [hep-th]].

\bibitem{Pius:2013tla} 
  R.~Pius and A.~Sen,
  ``S-duality Improved Perturbation Theory in Compactified Type I / Heterotic String Theory,''
  arXiv:1310.4593 [hep-th].

\end{thebibliography}
\end{document}